\documentclass[12pt]{article}

\usepackage[pdfpagelabels]{hyperref}

\usepackage[usenames,dvipsnames]{color}
\usepackage{graphicx,epsfig,xcolor}

\hypersetup{colorlinks=true, linkcolor=violet, urlcolor=blue, citecolor=blue}

\usepackage{cite}
\usepackage[english]{babel}
\usepackage{amssymb,amsfonts,amsmath}
\usepackage{verbatim}
\usepackage{bbm,bbold,bm}
\usepackage{tensor}
\usepackage{slashed}
\usepackage{braket}

\usepackage{comment}

\newcommand{\cE}{{\cal E}}  \newcommand{\cF}{{\cal F}}
\newcommand{\cG}{{\cal G}}  
  
  \newcommand{\cL}{{\cal L}}
\newcommand{\cM}{{\cal M}}  
  \newcommand{\cP}{{\cal P}}
  
  \newcommand{\cT}{{\cal T}}
  
\newcommand{\cW}{{\cal W}}

\def\q{\theta}
\def\a{\alpha}
\def\b{\beta}
\def\s{\sigma}
\def\m{\mu}
\def\n{\nu}
\def\r{\rho}

\def\be{\begin{equation}}
\def\ee{\end{equation}}
\def\ba{\begin{eqnarray}}
\def\ea{\end{eqnarray}}
\def\nb{\nonumber}
\def\p{\partial}

\def\dc{\nabla}  
 
\def\a{\alpha}
\def\b{\beta}

\def\e{\epsilon}

\def\g{\gamma}
\def\G{\Gamma}
\def\d{\delta}

\def\l{\lambda}

\def\m{\mu}
\def\n{\nu}

\def\r{\rho}
\def\s{\sigma}

\def\q{\quad}

\def\dd{\mathrm{d}}

\def\mc{\mathcal}

\newcommand{\pr}[1]{\left(#1\right)}
\newcommand{\pq}[1]{\left[#1\right]}
\newcommand{\pg}[1]{\left\{#1\right\}}

\newcommand{\one}[1]{#1^{(1)}}
\newcommand{\zero}[1]{#1^{(0)}}
\newcommand{\onef}[1]{\mc{#1}^{(1)}}
\newcommand{\zerof}[1]{\mc{#1}^{(0)}}

\usepackage{chngcntr}
\counterwithin*{equation}{section}

\usepackage{pifont}

\title{\bf Fracton Gauge Theories \\ in Curved Spacetimes}

\author{Evangelos Afxonidis${}^1$\footnote{afxonidisevangelos@uniovi.es}, Alessio Caddeo${}^2$\footnote{alessio.caddeo@uni-wuerzburg.de}, Carlos Hoyos${}^1$\footnote{hoyoscarlos@uniovi.es} and Daniele Musso${}^1$\footnote{mussodaniele@uniovi.es}
}

\date{}

\begin{document}
\maketitle 
\vspace{-20pt}
\begin{center}
\begin{minipage}{0.85\textwidth}
\begin{center}
\it{\small 
${}^1$Department of Physics, Universidad de Oviedo, and \\ Instituto de Ciencias and Tecnolog\'{\i}as Espaciales de Asturias (ICTEA)}\\  c/ Leopoldo Calvo Sotelo 18, ES-33007 Oviedo, Spain\\
\vspace{2mm}
${}^2$Institute for Theoretical Physics and Astrophysics and Würzburg-Dresden Cluster of Excellence ct.qmat 
Julius-Maximilians-Universität \\ Am Hubland, D-97074 Würzburg, Germany
\end{center}
\end{minipage}
\end{center}
\vspace{15pt}

\begin{abstract}
Fractonic matter with dipole symmetry can be coupled to a  two-index symmetric tensor gauge field. In this work, we show that this symmetric tensor field, along with other related generalized Maxwell theories, can be consistently coupled to curved backgrounds in a covariant and gauge-invariant way by reformulating dipole symmetry using conventional vector gauge fields. We identify a family of curved geometries where global dipole symmetry is preserved and derive energy-momentum conservation laws as Ward identities associated with background diffeomorphisms. Our results pave the way for future extensions, including generalizations to higher-order multipole theories.
\end{abstract}

\newpage

\tableofcontents

\vspace{10mm}

\section{Introduction}
\label{sec:intro}

Generalizations of Maxwell theory incorporating symmetric tensor fields were introduced by Pretko in \cite{Pretko:2016lgv}, motivated by their role in gapless spin liquids \cite{rasmussen2016stablegaplessboseliquid,Pretko:2016kxt}. These fields naturally couple to fractons—excitations with restricted mobility due to the conservation of the dipole moment or higher multipole moments of the charge. Fractons appear in various physical systems, including gapped excitations in lattice models designed for robust quantum memories, as well as in elastic and superfluid defects (see \cite{Nandkishore:2018sel,Pretko:2020cko,Grosvenor:2021hkn} for reviews). Symmetric tensor fields thus mediate interactions between fractons. In elasticity, for example, such fields emerge from a generalized version of particle-vortex duality \cite{Pretko_2018}, which describes the deformation of an elastic medium in the presence of disclinations or dislocations.

Coupling effective field theories to curved backgrounds has historically provided valuable insights, such as deriving conservation equations and identifying relations between transport coefficients (see, \emph{e.g.} \cite{Son:2005rv,Hoyos:2011ez,Son:2013rqa}). However, extending this approach to symmetric tensor theories has proven challenging. Existing attempts have succeeded only under special geometric conditions \cite{Slagle:2018kqf,Bidussi:2021nmp,Jain:2021ibh,Armas:2023ouk}, by introducing additional Stueckelberg fields \cite{Pena-Benitez:2021ipo,Jain:2024ngx}\footnote{These were also used to generalize the coupling of matter to dipole gauge fields in \cite{Afxonidis:2023tup}.}, or in theories that do not reduce to Pretko’s in flat spacetime \cite{Gromov:2017vir,Hartong:2024hvs}. Although elasticity can be consistently coupled to a curved geometry, the dualization of elastic displacement fields into symmetric tensor fields requires the metric itself to be dualized into higher-rank tensors \cite{Tsaloukidis:2023bvz}.

At the heart of this difficulty lies the nature of gauge transformations for symmetric tensor fields, which conflicts with general covariance. However, dipole symmetry can also be realized using ordinary one-form gauge fields, as demonstrated in \cite{Caddeo:2022ibe} in the context of reformulating the particle-vortex dual of elasticity. In the present work, we show that, using such an alternative formulation for the dipole gauge symmetry, Pretko’s theory can be consistently coupled to an arbitrary curved (Aristotelian) background in a manifestly covariant way. While {\em global} dipole symmetry is generally broken in a curved background, we identify special nontrivial cases where it remains preserved, in particular when time and space factorize and the spatial manifold is conformally flat. Furthermore, we derive energy-momentum conservation laws as Ward identities of background diffeomorphisms and demonstrate that these identities remain gauge-invariant on-shell, allowing them to be formulated in terms of gauge-invariant currents. This is in contrast to other formulations, where the momentum density should shift under a dipole transformation.

The structure of the paper is as follows. In Section~\ref{sec:gaugeingredients}, we introduce the formalism and establish the connection to Pretko’s theory in flat spacetime.  
In Section~\ref{sec:curved}, we develop the coupling of dipole gauge fields to curved backgrounds, derive the conditions for global dipole symmetry, and extract the Ward identities associated with diffeomorphisms.  
In Section~\ref{sec:pretko}, we construct the mapping to the symmetric tensor theory in a curved background and discuss energy and momentum conservation in flat space.  
We conclude with a discussion of our results in Section~\ref{sec:discuss}.  
Additional technical details are provided in the appendices.  

\section{Gauge fields for dipole symmetry}
\label{sec:gaugeingredients}

Let us review the formalism introduced in \cite{Caddeo:2022ibe} to write fractonic theories using ordinary gauge fields.
We work in generic $d+1$ spacetime dimensions.
The generators of internal symmetries are internal translations $P_a$, Abelian $U(1)$ monopole charge symmetry $Q$, and internal dipole transformations $Q^a$. They span a Monopole-Dipole-Moment Algebra (MDMA), with the only non-trivial commutator being
\begin{equation}
i[P_a,Q^b]=\delta_a^b Q \ .
\end{equation}
Here, $a,b=1,\dots, d$ are internal indices, and as such do not transform under spacetime symmetries. Spacetime indices will be denoted by Greek letters, and purely spatial by mid-alphabet latin letters $i,j,k$, etc. Note that these generators are independent of the analogous spacetime translations and dipole transformations and do not act on the coordinates.

We introduce a gauge potential
\begin{equation}
{\cal A}_\mu= V_\mu^{a}P_a+a_\mu Q+ b_{\mu a} Q^a\ .
\end{equation}
Since the index $a$ is internal, the gauge fields $V_{\m}^{a}$, $a_{\m}$ and $b_{\m a}$ are the components of one-form fields.
An infinitesimal gauge transformation takes the usual form
\begin{equation}
\delta_\Lambda {\cal A}_\mu =D_\mu \Lambda=\partial_\mu \Lambda+i[{\cal A}_\mu,\Lambda]\ ,
\end{equation}
where the gauge parameter, expanded on the internal symmetry generators, is parameterized as
\begin{equation}
\Lambda=\kappa^a P_a+\lambda Q+\lambda_{ a} Q^a \ .
\end{equation}
In this paper, we will often use the differential forms notation, which allows one to easily understand how an object depends on the spacetime metric. In this language, denoting $\dd$ as the exterior derivative, the gauge field transformations read
\begin{subequations}
\ba 
\delta V^{a} &=& \dd \kappa^a \  ,\\
\delta a &=& \dd \lambda+V ^{ a}\lambda_{a}-b_{ a}\kappa^a \ ,\\
\delta b_{ a}&=& \dd \lambda_{a} \ .
\ea
\end{subequations}
Defining $f \equiv \dd a$, we have two curvature invariants
\begin{subequations}
\ba
\mc{V} ^a &=& \dd  V ^{a}  \ ,\\
\mc{F}^{(1)}_{a} &=& \dd  b_{ a}  \ ,
\ea
\end{subequations}
and a  curvature
\begin{equation}
\label{defBmunu}
\mc{F}^{(0)}  = f  - b_{ a} \wedge  V ^{ a} \ ,
\end{equation}
transforming as
\begin{equation}
\label{eq:gaugetransfcurvature2}
    \delta \mc{F}^{(0)} 
    =
     \mc{V} ^a \lambda_{ a}
    - \mc{F}^{(1)}_{ a}\kappa^a\ .
\end{equation}
To formulate fractonic gauge theories, we treat $a _{\m}$ and $b_{\m a}$ as dynamical fields, whereas $V_{\m}^{a}$ is taken as a background field. In Section \ref{sec:mapflat}, we see that the relevant generalized Maxwell theories, such as Pretko's model, can be obtained using actions that involve the curvature $\zerof{F}$. From (\ref{eq:gaugetransfcurvature2}) we understand that requiring dipole invariance of $\zerof{F}$ imposes the condition
\be\label{eq:zerocurvV}
\mc{V}^{a} = \dd V^{a} = 0 \ .
\ee
Furthermore, internal translations are explicitly broken by $\zerof{F}$.

\subsection{Dipole current conservation equations}
\label{sec:dipolewardflat}

Gauge invariance under monopole and dipole transformations is encoded in Ward identities. 
Let us consider the coupling of the dynamical fields to external currents, 
\be
S_{J}= \int d^{d+1} x \pr{   a_{\m} J^{\m} + b_{\m a} J^{\m a} }\ .
\ee
Under a gauge transformation,
\ba
\d S_{J} &=& \int d^{d+1} x \pr{  \p_{\m} \l J^{\m} + \l_{a}  \d^{a}_{\m} J^{\m} + \p_{\m} \l_{ a} J^{\m a}} \nb \\
&=& \int d^{d+1} x \pq{- \l \p_{\m} J^{\m} +  \l_{a} \pr{ \d^{a}_{\m} J^{\m} - \p_{\m}  J^{\m a}}} \ , 
\ea
where we integrated by parts assuming no boundary contributions. We thus obtain the Ward identities for the currents
\be
\p_{\m} J^{\m a} =  \d_{\m} ^{a} J^{\m} \ , \q \q \q \p_{\m} J^{\m} = 0 \ .
\ee
The currents are defined up to the following improvements,
\begin{subequations}
\ba
J^{\m} &\rightarrow& J^{\m} + \p_{\n} S^{\m \n}_{_M} \ ,  \\ 
J^{\m a}  &\rightarrow& J^{\m a} + \p_{\n} S^{\m \n a} _{_D} -  \d_{\n}^{a} S_{_M} ^{\m \n} \ ,
\ea
\end{subequations}
where $S^{\m \n}_{_M} = - S^{\n \m}_{_M} $ and $S^{\m \n a}_{_D} = - S^{\n \m a}_{_D}$.
Deriving the first Ward identity, we find
\be
\p_{0} \p_{i} J^{0i} + \p_{i} \p_{j} J^{ij} =  \p_{i} J^{i} = -  \p_{0} J^{0} \ .
\ee
We can choose improvement terms so that $J^{0i} = 0$ and $J^{ia}=J^{ai}$, in such a way that
\be
\p_{0} J^{0} + \p_{i} \p_{j} J^{ij} = 0 \ ,
\ee
which is the usual conservation law for theories with dipole symmetry with a symmetric tensor current. This is how we retrieve the usual conservation law of fracton gauge theories starting from our formulation.

\subsection{Generalized Maxwell theories and Pretko's model}\label{sec:mapflat}

In this section, we work in flat spacetime and show how generalized Maxwell theories, and in particular Pretko's model, are obtained using our formalism. Section \ref{sec:curved} is instead dedicated to the coupling of these models to curved geometries.

In the flat-spacetime case, the background field $V_{\m}^{a}$ can (and will) be taken as
\be
V_{\m}^{a} = \d^{a}_{\m} \ .
\ee
On such a background, internal indices and (spatial) spacetimes indices can be traded with each other. As a consequence, the internal transformations will generate global spacetime symmetries, particularly spatial dipole symmetry, as we will show.

The gauge-invariant curvatures become
\ba
\mc{F}^{(1)}_{\mu\nu a} &=& \p_\mu \pr{b_{\nu a}- \p_{\n} a_{a}}-\p_\nu \pr{b_{\mu a} - \p_{\m} a_{a}} \ , \\
\mc{F}^{(0)}_{\m \n} 
&=&  - \pr{ b_{\m a} \d_{\n}^{a} - \p_{\m} a_{\n}} + \pr{ b_{\n a} \d_{\m}^{a} - \p_{\n} a_{\m}} \ ,
\ea
and satisfy the following Bianchi identity 
\be
\label{eq:bianchi_identities}
 \e^{\m \n \r \s} \p_{\r} \mc{F}^{(1)}_{\m \n a} = 0 \ .
\ee
We can construct combinations corresponding to field strengths of different possible theories
\begin{subequations}\label{eq:bianchifields}
\ba 
F_{0ia} &=& \a \onef{F}_{0ia} - \b \,2\partial_{[0}\zerof{F}_{i]a} - \zeta \p_{0} \zerof{F}_{ia} \ , \\ 
F_{ija} &=& \a \onef{F}_{ija} -(\b+\zeta)\, 2 \partial_{[i}\zerof{F}_{j]a} \ , 
\ea
\end{subequations}
where $\alpha$, $\beta$ and $\zeta$ are constant coefficients.
One can check that the Bianchi identity is satisfied for all these combinations:
\begin{equation}
   \epsilon^{\mu\nu\rho\sigma}\partial_\rho F_{\mu\nu a}=0 \ .
\end{equation}
For the special case $\a=-\b=2 \zeta$, the field strength depends only on the symmetric components of the gauge field. 
Without loss of generality, we can set $\a =1$ for this case. Defining $\phi \equiv a_{0}$ and the symmetric tensor potential 
\be
\label{eq:defsymmetricfieldinflatcase}
A_{ij} \equiv 2 \pr{b_{(ij)} - \p_{(i} a_{j)}} \ ,
\ee
the field strength of the symmetric tensor theory takes the form introduced by Pretko: 
\begin{equation}
\label{eq:fieldstrengthPretkoflat}
    F_{0i a}=-F_{i0 a} = F_{0ai} =\frac{1}{2}\partial_0 A_{ia}+\partial_i \partial_a \phi \ , \q \q \q  F_{ij a}=\partial_{[i}A_{j]a} \ .
\end{equation}
We notice in particular that for this choice of parameters, the electric field $F_{0ia}$ is symmetric. The gauge transformations of the potentials also reproduce those studied by Pretko, namely
\be
\label{eq:flatnewfieldsgaugetransformations}
\d A_{ij} = - 2 \p_{i} \p_{j} \l \ , 
\q \q \d \phi = \p_{0} \l \ .
\ee
Any transformation of parameter $\lambda(x)$ that leaves the gauge potentials invariant can be identified with a global symmetry transformation. Under such a transformation, and in the absence of matter fields, the theory is trivially invariant. For constant parameters $c$ and $d_i$, the global symmetry transformations correspond to
\begin{equation}\label{eq:globalgauge}
    \lambda(x)=c+d_i x^i\,.
\end{equation}
A field $\Phi$ of Abelian charge $q$ would transform as
\begin{equation}
    \Phi(x)\to e^{i q \lambda}\Phi(x) =e^{i q c}e^{i q d_i x^i}\Phi(x)\,.
\end{equation}
Thus, $c$ parametrizes $U(1)$ monopole transformations while $d_i$ parametrizes spatial dipole transformations.

Let us now consider the simple generalized Maxwell action
\be
\label{eq:FractonMaxwellLagrangian}
S=\frac{1}{2}\int d^{d+1}x\pr{F_{0ia}F_{0ia}-\frac{\s}{2} F_{kla}F_{kla}} \ ,
\ee
where $\sigma$ is a constant parameter that can be identified with the squared of the speed of sound.
Since these field strengths are always built with the derivative of the gauge fields, the spectrum features only massless excitations.%
\footnote{In order to introduce a mass gap, we would need to add to the Lagrangian terms like $\mc{F}^{(0)}_{ij}\mc{F}^{(0)}_{ij}$ or $\mc{F}^{(0)}_{0j}\mc{F}^{(0)}_{0j}$, but we refrain from doing it.}

From the variation of the action with respect to $b_{\mu a}$ and $a_\mu$, one obtains the equations of motion. Including also the Bianchi identities, the generalized Maxwell's equations for the symmetric tensor theory are
\begin{subequations}
\label{eq:eomssymm}
\begin{align}
  &\partial_0 F_{0ia}-\sigma \partial_k F_{k(ia)}=0 \ ,\  &\partial_i \partial_j F_{0ij}=0 \ ,\\
  & \partial_0 F_{ij a}+2\partial_{[i}F_{j]0a}=0 \ ,\ &\partial_{[i}F_{jk]a}=0 \ .
\end{align}
\end{subequations}
These equations map to ordinary Maxwell's equations for a field strength defined by taking a derivative along the internal index, namely $\tilde{F}_{\mu\nu}\equiv\partial_l F_{\mu\nu l}$:
\begin{align}\label{eq:tildeeomssymm}
  &\partial_0 \tilde{F}_{0i}-\frac{\sigma}{2} \partial_k \tilde{F}_{ki}=0 \ ,\  &\partial_i  \tilde{F}_{0i}=0 \ ,\\
  & \partial_0 \tilde{F}_{ij}+2\partial_{[i}\tilde{F}_{j]0}=0 \ ,\ &\partial_{[i}\tilde{F}_{jk]}=0 \ .
\end{align}
We will use these equations to check the diffeomorphisms Ward identities in Section \ref{sec:flatenergystress}.

Since in general we deal with non-relativistic theories, it is sometimes convenient to introduce electric and magnetic fields separately. For simplicity, and because it is the case that is more frequently studied, we just give the expressions for  $d=3$ spatial dimensions
\begin{equation}
    E_{ia}=F_{0ia} \ ,\q \q \q  B_{ia}=\frac{1}{2}\epsilon_{ikl}F_{kla} \ .
\end{equation}
The action and the equations of motion in terms of the electric and magnetic fields are respectively
\begin{equation}
    S=\frac{1}{2}\int d^{3+1}x\,  \pr{E_{ia} E_{ia} - \s B_{ia} B_{ia}} \ ,
\end{equation}
and 
\begin{align}
   &\partial_0 E_{ia}+\sigma\epsilon_{kl(i|}\partial_k  B_{l|a)}=0 \ ,\  &\partial_k \partial_l E_{kl}=0 \ ,\\
    &\partial_0  B_{ia}-\epsilon_{kli}\partial_k  E_{la}=0 \ ,\ &\partial_k B_{ka}=0 \ .
\end{align}
We recognize the second identity in the first equation as the generalized Gauss' law for a theory with conserved dipole moment \cite{Pretko:2016lgv}.

\section{Fracton gauge theories in curved spacetimes}\label{sec:curved}

We now want to present a curved spacetime formulation of the generalized Maxwell theories discussed in Section \ref{sec:mapflat}. Since fracton models are non-relativistic, we consider Newton-Cartan geometries \cite{Son:2013rqa,Geracie:2014nka,Brauner:2014jaa,Jensen:2014aia}, which, in the absence of boost invariance, are also known as Aristotelian geometries. As a result, besides the background field $V_{\m}^{a}$, our geometrical background will be characterized by a clock one-form $n_{\m}$, a velocity field $v^{\m}$, and the degenerate covariant and contravariant metrics $h_{\m \n}$, $h^{\m \n}$. These fields satisfy the conditions
\begin{equation}\label{eq:NCconstraints}
    v^\mu n_\mu=1 \ ,\qquad v^\mu h_{\mu\nu}=0 \ ,\qquad n_\mu h^{\mu\nu}=0 \ ,\qquad h^{\mu\alpha} h_{\alpha\nu}=P^\mu_\nu\equiv \delta^\mu_\nu-v^\mu n_\nu \ .
\end{equation}
The clock form and velocity field project on a time-like direction while the degenerate metrics project on space-like directions. It is also common to introduce the non-degenerate spacetime metric $\gamma_{\mu\nu}=n_\mu n_\nu+h_{\mu\nu}$ and its inverse $\gamma^{\mu\nu}=v^\mu v^\nu+h^{\mu\nu}$. In Appendix \ref{app:NewtonCartanreview} we present some additional facts of Newton-Cartan geometries. Moreover, we will often employ differential forms. Not only do they yield more concise formulae, but they are also automatically covariant. As a result, any identity expressed in differential form notation will hold in any spacetime.

In Section \ref{sec:gaugeingredients}, we presented the gauge fields $a_{\m}$ and $b_{\m a}$, their gauge transformations and the curvatures $\zerof{F}$, $\onef{F}$ in terms of differential forms. Therefore, these definitions do not require any generalization.  Regarding the background field $V_{\m}^{a}$, in flat-spacetime we chose $V_{\m}^{a} = \d^{a} _{\m}$, linking internal and spatial indices. In the curved case, instead, we just require 
\be
\label{eq:closure_V}
i_{v} V^{a}=0  \ ,
\ee
where $i$ indicates the interior product on forms, $i_{v} V^{a}=v^{\m} V_{\m}^{a}$. This condition removes time-like components from $V^a$, which acts in many cases as a frame field with internal and space-like spacetime indices. In addition, we keep the condition that $V^a$ is a closed form \eqref{eq:zerocurvV}. 

In Section \ref{sec:mapflat}, we saw that the models that we study in this paper are formulated in terms of a field strength constructed from the curvature $\onef{F}$ and derivatives of $\zerof{F}$. as in (\ref{eq:bianchifields}). 
The curvature $\onef{F}$ is a two-form with an internal index; we thus want to built from derivatives of $\zerof{F}$ objects with the same index structure. To this purpose, we first introduce
\be
V^{\m}_a =\delta_{ab} h^{\m \n} V_{\n}^{b} \ , \q \q \q n_{\m} V^{\m}_a =0 \ ,
\ee
and then define
\be
\label{eq:defcurvaturesG}
\zero{\mc{G}}_{a} = \dd \, i_{V^{a}/V^{2}}  \zero{\mc{F}}  \ , \q \q 
\zero{\tilde{\mc{G}}}_{a} = \dd \, i_{V^{a}/V^{2}} \pr{ \zero{\mc{F}} - n \wedge i_{v} \zero{\mc{F}} } \ , 
\ee
where, recalling that $d$ is the number of spatial dimensions, 
\be
\label{eq:defV2}
V^{2} \equiv \frac{1}{d} V_{\m}^{a} V^{\m}_{a} \ .
\ee
It is immediate to see that all the curvatures so defined satisfy the Bianchi identity,
\be
\label{eq:bianchicurved}
\dd \onef{F} = \dd  \zero{\mc{G}}_{a} = \dd \zero{\tilde{\mc{G}}}_{a} = 0 \ .
\ee

From the field strengths, we can define the electric fields
\be
\one{E}_{a} = i_{v} \one{\mc{F}}_{a} \ , \q \q \q \zero{E}_{a} = i_{v} \zero{\mc{G}}_{a} \ , \q \q \q \zero{\tilde{E}}_{a} = i_{v} \zero{\tilde{\mc{G}}}_{a} \ .
\ee
These are one-forms in spacetime, with vanishing temporal component because $i_{v}^{2}=0$. The magnetic fields are defined through (see Appendix \ref{app:NewtonCartanreview} for details on the definition of Hodge dual)
\be
\one{B}_{a} = i_{v} * \one{\mc{F}}_{a} \ , \q \q \q \zero{B}_{a} = i_{v} * \zero{\mc{G}}_{a} \ , \q \q \q \zero{\tilde{B}}_{a} = i_{v} * \zero{\tilde{\mc{G}}}_{a} \ .
\ee
These are $(d-2)$-forms with vanishing temporal components.
We can then combine them,\footnote{One can easily show that in the flat-spacetime case $v^{\m} = \d^{\m}_{0}$, $n_{\m}= \d_{\m}^{0}$, $h^{\m 0 }=0$, $h^{ij}=\d^{ij}$, $V_{\m}^{a}=\d^{a}_{\m}$, these combinations reduce to those in (\ref{eq:bianchifields}).}
\begin{subequations}
\label{eq:bianchifieldscurved}
\ba
E_{a} &=& \a E_{a}^{(1)} + \b E_{a}^{(0)} + \zeta \tilde{E}_{a}^{(0)} \ , \\ 
B_{a} &=& \a B_{a}^{(1)} + \b B_{a}^{(0)} + \zeta \tilde{B}_{a}^{(0)}  \ .
\ea
\end{subequations}
Once we have the electric and magnetic fields, we can write the action
\be
\label{eq:MaxwellLagrangianwithforms}
S = \frac{1}{2} \int \pr{E_{a} \wedge * E_{a} - \s B_{a} \wedge * B_{a}} \ , 
\ee
with a sum over the internal indices. This action is gauge and diffeomorphism invariant. In the following, it will be useful to 
rewrite (\ref{eq:MaxwellLagrangianwithforms}) also in components as
\begin{equation}\label{eq:F2}
    S=\int d^{d+1} x\, G^{\mu\nu a}F_{\mu\nu a} \ ,
\end{equation}
where we have introduced the field strength
\begin{equation}
F_{\mu\nu a}=\alpha\cF_{\mu\nu a}^{(1)}-2\partial_{[\mu}\left[(\beta+\zeta) \cF^{(0)}_{\nu]\lambda}\frac{V_a^\lambda}{V^2} -\zeta n_{\nu]} \cF^{(0)}_{\rho\lambda}v^\rho \frac{V_a^\lambda}{V^2} \right] \ ,
\end{equation}
and defined for later convenience ($\g = \text{det}( \g_{\m \n})$)
\begin{equation}
G^{\mu\nu a}=\sqrt{\gamma}\, \frac{\delta^{ab}}{2}\left(v^{[\mu} h^{\nu]\sigma}v^\rho -\frac{\sigma}{2} h^{\mu\rho}h^{\nu\sigma}\right)F_{\rho\sigma b} \ .
\end{equation}

\subsection{Ward identities for the monopole and dipole currents}

In Section \ref{sec:dipolewardflat}, we studied the Ward identities for dipole conservation in flat spacetime. 
Let us consider the general curved case. Coupling our gauge fields to external currents, the action $S$ acquires the term 
\be\label{eq:actioncurrents}
S_{J} = \int \pr{a \wedge * \zero{J} + b_{a} \wedge  * \one{J}_{a}} \ .
\ee
Under a gauge variation, and after integration by parts, we get
\be
\d S = \int \pr{- \l \wedge \dd  * \zero{J} +\l_{a} V^{a}\wedge   * \zero{J}  - \l_{a} \dd * \one{J}_{a}}\ ,
\ee
from which we obtain the Ward identities
\be
\label{eq:Ward_id}
\dd * \zero{J} = 0 \ , \q \q \q \dd * \one{J}_{a} =  V^{a} \wedge * \zero{J} \ .
\ee
In components these yield
\be
\label{eq:curvedWardnoncovariant}
\p_{\m} \pr{ \sqrt{\g}  J^{\m}} = 0\ , \q \q \q 
\frac{1}{\sqrt{\g} } \p_{\m} \pr{\sqrt{\g}  J^{\m a}} = V^{a}_{\r} J^{\r}  \ ,
\ee
or, in terms of the Newton-Cartan covariant derivative, 
\be
\pr{\dc_{\m} - T^{\n}_{\n \m}}   J^{\m} = 0\ , \q \q \q 
\pr{ \dc_{\m} - T^{\n}_{\n \m}}    J^{\m a} = V^{a}_{\r} J^{\r}  \ ,
\ee
where $T^\lambda_{\mu\nu}$ is the torsion tensor (see Appendix \ref{app:NewtonCartanreview}).

Let us introduce the improved currents
\begin{equation}
    \label{eq:impro}
     \tilde J^\mu=J^\mu-\frac{1}{\sqrt{\gamma}}\partial_\lambda(\sqrt{\gamma}S^{\lambda\mu}) \ , \q \quad \tilde J^{\mu a}=J^{\mu a}- S^{\mu\nu}V_\nu^a\ ,
\end{equation}
with $S^{\mu\nu}=-S^{\nu\mu}$. Using that $\partial_{[\mu} V_{\nu]}^a=0$, one can check that they satisfy the same Ward identities \eqref{eq:Ward_id} as the original currents
\begin{equation}
     \partial_\mu(\sqrt{\gamma}\tilde J^\mu)=0 \ , \q \q \q \frac{1}{\sqrt{\gamma}}\partial_\mu(\sqrt{\gamma}\tilde J^{\mu a})=V_\mu^a \tilde J^\mu \ .
\end{equation}
We can use the improvement \eqref{eq:impro} to fix
\begin{equation}
    \label{eq:Ward_improved}
    n_\mu \tilde J^{\mu a}=0 \ , \q \q \q V_\mu^a \tilde J^{\mu b}-V_\mu^b \tilde J^{\mu a}=0 \ .
\end{equation}
by taking
\begin{equation}
    \label{eq:S_impro}
    S^{\mu\nu}=2 \frac{v^{[\mu}V^{\nu]}_a}{V^2} n_\lambda J^{\lambda a}
    -\frac{V_a^{[\mu}P^{\nu]}_{\ \lambda}}{V^2}  J^{\lambda a} \ ,
\end{equation}
and using
\begin{equation}
    P^\mu_{\ \nu}=\frac{V^\mu_a V^a_\nu}{V^2} \ , \qquad 
    V_\mu^a V^\mu_c = V^2 \delta^a_c\ .
\end{equation}
We can combine the Ward identities \eqref{eq:curvedWardnoncovariant} for such improved currents into the dipole current conservation equation
\begin{equation}\label{eq:wardcurved}
    \partial_\mu(\sqrt{\gamma}v^\mu n_\lambda \tilde J^\lambda)+\partial_\mu\left[\frac{V^\mu_a}{V^2}\partial_\lambda(\sqrt{\gamma} \tilde J^{\lambda a}) \right]=0 \ .
\end{equation}

We will use the conservation equation to study the conditions for having a global dipole symmetry. In flat spacetime, we can write a current depending on a scalar function $\chi$
\begin{equation}
    J_\chi^{\mu}=\tilde J^\mu \chi-\tilde J^{\mu a}\delta_a^i\partial_i \chi \ . 
\end{equation}
The divergence of this current is
\begin{equation}
    \partial_\mu J_\chi^{\mu}=\tilde J^\mu \partial_\mu \chi- \tilde J^i\partial_i \chi-\tilde J^{\mu a}\delta_a^i \partial_\mu \partial_i \chi=\tilde J^0 \partial_0 \chi-\tilde J^{ij}\partial_i\partial_j \chi =0 \ . 
\end{equation}
We get two conditions
\begin{equation}
    \partial_0 \chi=0 \ , \q \q \quad \partial_i \partial_j \chi=0 \ ,
\end{equation}
with solutions $\chi=c+d_i x^i$, where the coefficients $c,d_i$ are constant and the first corresponds to the monopole and the second to the dipole conservation. Note that the solution for $\chi$ coincides precisely with the global symmetry transformations \eqref{eq:globalgauge}, and $J_\chi^\mu$ is the conserved matter current associated to this symmetry. 

We can generalize this to curved spacetime, using a definition of the current
\begin{equation}
    J_\chi^{\mu}=\tilde J^\mu \chi-\tilde J^{\mu a}\frac{V_a^\lambda}{V^2}\partial_\lambda \chi \ . 
\end{equation}
The covariant derivative is
\ba
  0&=&
  \frac{1}{\sqrt{\gamma}} \partial_\mu (\sqrt{\gamma} J^{\mu}_\chi) =\tilde J^\mu \partial_\mu \chi-\frac{V_\mu^a  V_a^\lambda}{V^2}\tilde J^\mu\partial_\lambda \chi-\tilde J^{\mu a}\partial_\mu\left(\frac{V_a^\lambda}{V^2}\partial_\lambda \chi\right) \nb \\
  &=&n_\mu \tilde J^\mu v^\lambda\partial_\lambda\chi-\tilde J^{\mu a}\partial_\mu\left(\frac{V_a^\lambda}{V^2}\partial_\lambda \chi\right) \ .
\ea
We arrive at the conditions
\begin{equation}
    v^\lambda \partial_\lambda \chi=0 \ , \q \q \q  V_{(b}^\mu\partial_\mu\left(\frac{V_{a)}^\lambda}{V^2}\partial_\lambda \chi\right)=0 \ ,
\end{equation}
where we have used the second relation in \eqref{eq:Ward_improved}.
There is always a solution $\chi=c$ corresponding to the conservation of the monopole charge. Other possible solutions for $\chi$, if they exist, would correspond to the curved space analog of a conserved dipole charge.

Examples are spaces $\mathbb{R}_t\times {\cal M}_d$ where ${\cal M}_d$ is conformally flat
\begin{equation}
    n_\mu=\delta_\mu^0 \ ,
    \qquad v^\mu=\delta^\mu_0 \ ,
    \qquad h_{0\mu}=h^{0\mu}=0 \ ,
    \qquad h_{ij}=e^{2\omega}\delta_{ij} \ ,
    \qquad h^{ij}=e^{-2\omega}\delta^{ij} \ .
\end{equation}
Fixing $V_0^a=0,\ V_i^a=\delta_i^a$, we get the same equations for $\chi$ as in flat spacetime, thus we have dipole conservation in this type of spacetimes. That includes all $d=2$ manifolds and spaces with vanishing Cotton tensor ($d=3$) or Weyl tensor ($d\geq 4$).
\subsection{Ward identity for diffeomorphisms}

After the gauge fields have been integrated out, one is left with a generating functional $\cW[n_\mu,v^\mu,h^{\mu\nu},h_{\mu\nu},V_\mu^a]$ depending on the background fields. These transform as tensors under background diffeomorphisms. Denoting the Lie derivative by $\cL_\xi$, one has
\begin{subequations}
 \label{eq:vars}
\ba
    \delta_\xi n_\mu &=& \cL_\xi n_\mu \ ,\q \q \q 
     \delta_\xi v^\mu=\cL_\xi v^\mu \ ,\\
     \delta_\xi h^{\mu\nu}&=& \cL_\xi h^{\mu\nu} \ ,\q \q  \ 
    \delta_\xi V_\mu^a=\cL_\xi V_\mu^a=\partial_\mu(\xi^\alpha V_\alpha^a)\ ,
\ea
\end{subequations}
where we have used that $\xi^\alpha\partial_\alpha V_\mu^a=\xi^\alpha\partial_\mu V_\alpha^a$, a direct consequence of the closure of $V^a$ \eqref{eq:zerocurvV}.

The variation of the generating functional is
\begin{equation}\label{eq:genfuncvar}
    \delta_\xi \cW=\int d^{d+1}x\,\sqrt{\gamma}\left(\cM_a^\mu \partial_\mu(\xi^\alpha V_\alpha^a) -\cE^\mu \cL_\xi n_\mu-\cP_\mu\cL_\xi v^\mu-\frac{1}{2}\cT_{\mu\nu}\cL_\xi h^{\mu\nu}-\frac{1}{2}\widetilde \cT^{\mu\nu}\cL_\xi h_{\mu\nu}\right) \ .
\end{equation}
However, not all the variations are independent, due to the constraints \eqref{eq:NCconstraints} and \eqref{eq:closure_V}. Denoting from now on the unconstrained variations by means of a bar, for a general variation of the background fields, we have
(see \cite{Jensen:2014aia})
\begin{subequations}
\ba
\d n_\m &=& \d \bar n_\m \ , \\
\d v^\m &=& - v^\m v^\n \d \bar n_\n + P^\m _\n \d \bar v^\n \ , \\
\d h^{\m \n} &=& - \pr{v^\m h^{\n \r} + v^\n h^{\m \r}} \d \bar n_\r + P^\m _\r P^\n _\s \d \bar h^{\r \s} \ , \\
\d h_{\m \n} &=& - \pr{n_\m h_{\n \r} + n_\n h_{\m \r}} \d \bar v^\r - h_{\m \r} h_{\n \s} \d \bar h^{\r \s} \ , \\
\d V_\m^a &=& -n_\mu \d \bar v^\r V_\r^a+ P_\m^\r \d \bar V_\r^a \ ,
\ea
\end{subequations}
From these and recalling \eqref{eq:vars}, we obtain
\begin{subequations}
\ba
n_\m \cL_\xi v^\m &=& - v^\mu\cL_\xi n_\mu \ ,\\
n_\m n_\n \cL_\xi h^{\m \n} &=& 0 \ ,\q \quad
n_\m P_\n^\r \cL_\xi h^{\m\n} = -h^{\r\m}\cL_\xi n_\m\ ,\\
v^\m v^\n \cL_\xi h_{\m\n} &=& 0 \ ,\q \quad
v^\m P^\n_\r \cL_\xi h_{\m\n} = - h_{\r\m}\cL_\xi v^\m \ ,\\
P_\r^\m P_\s^\n \cL_\xi h_{\m\n} &=& -h_{\r\m}h_{\r\s}\cL_\xi h^{\m\n} \ ,\\
v^\m\partial_\m (\xi^\a V_\a^a) &=& -V_\m^a \cL_\xi v^\m \ .
\ea
\end{subequations}
After integrating by parts the first term in \eqref{eq:genfuncvar}, we use these relations (except the last one) to rewrite the variation of the generating functional as
\begin{equation}
    \delta_\xi \cW=-\int d^{d+1}x\,\sqrt{\gamma}\left(\frac{1}{\sqrt{\gamma}}\partial_\mu(\sqrt{\gamma}\cM_a^\mu) \xi^\alpha V_\alpha^a +\bar\cE^\mu \cL_\xi n_\mu +\bar\cP_\mu\cL_\xi v^\mu +\frac{1}{2}\bar\cT_{\mu\nu}\cL_\xi h^{\mu\nu}\right) \ ,
\end{equation}
where
\ba
\overline\cE^\m &=& \cE^\m-v^\m v^\r\cP_\r-\frac{1}{2}(h^{\m\r} v^\s+h^{\m\s} v^\r) \cT_{\r\s} \ ,\\
\overline \cP_\m &=&  P_\m^\r \cP_\r-\frac{1}{2}(h_{\m\r} n_\s+h_{\m\s} n_\r) \widetilde \cT^{\r\s} \ ,\\
\overline \cT_{\m\n} &=&  P_\m^\r  P_\n^\s \cT_{\r\s}-h_{\m\r}h_{\n\s}\widetilde\cT^{\r\s} \ .
\ea
These can be identified as the energy, momentum and stress currents respectively. Note that 
\begin{equation}
    \label{eq:trans_def}
    v^\m \overline \cP_\m= 0=v^\m \overline \cT_{\m\n}\ ,
\end{equation}
and we say they are thus \emph{transverse}. The conditions $\delta_\xi\cW=0$ gives the Ward identities for diffeomorphisms
\begin{align}\label{eq:wardid}
    &
    -\frac{1}{\sqrt{\gamma}}\partial_\mu(\sqrt{\gamma}\cM^\m_a)V_\a^a
    +
    \frac{1}{\sqrt{\gamma}}\partial_\mu\left[\sqrt{\gamma} (n_\a \overline \cE^\m -v^\m \overline \cP_\a -h^{\m \n}\overline{\mc{T}}_{\n\a})\right]
    \\ & \nonumber \qquad 
    -\overline \cE^\m\partial_\a n_\m-\overline  \cP_\m\partial_\a v^\m-\frac{1}{2}\overline \cT_{\m\n}\partial_\a h^{\m\n}=0\ ,
\end{align}
or, in covariant form,
\begin{align}\label{eq::WardIdentitiesdiffeos}
    &-\frac{1}{\sqrt{\gamma}}\partial_\mu(\sqrt{\gamma}\cM^\m_a)V_\a^a
    +
    n_{\alpha}\nabla_{\mu}\overline{\mathcal{E}}^{\mu}-\nabla_{\mu}\left( v^{\mu}\overline{\mathcal{P}}_{\alpha}\right)-h^{\mu\nu}\nabla_{\mu}\overline{\mathcal{T}}_{\nu\alpha}
    \\ \nonumber & 
    \qquad \qquad
    -T^{\lambda}_{\lambda\mu}\left( n_{\alpha}\overline{\mathcal{E}}^{\mu}-v^{\mu}\overline{\mathcal{P}}_{\alpha}-h^{\mu\nu}\overline{\mathcal{T}}_{\nu\alpha}\right)-\overline{\mathcal{P}}_{\mu}\nabla_{\alpha}v^{\mu}
    +T^\l_{\m \a} n_\l \overline{ \mc{E}}^\mu=0 \ ,
\end{align}
where we have used that $\overline{\mathcal{P}}_{\mu}$ and $\overline{\mathcal{T}}_{\mu\nu}$ are transverse. Projecting with $v^{\alpha}$ and $P^{\alpha}_{\beta}$, we obtain
\begin{align}
    &\left( \nabla_{\mu}-T^{\lambda}_{\lambda\mu}\right)\overline{\mathcal{E}}^{\mu}=-h^{\mu(\nu}\nabla_{\mu}v^{\alpha)}\overline{\mathcal{T}}_{\nu\alpha}-T^{\lambda}_{\mu\alpha}n_{\lambda}v^{\alpha}\overline{\mathcal{E}}^{\mu} \ , \\
    \nb \\
    &P^{\alpha}_{\beta}\left[\left( \nabla_{\mu}-T^{\lambda}_{\lambda\mu}\right)\left( v^{\mu} \overline{\mathcal{P}}_{\alpha}\right) +h^{\mu\nu}\left( \nabla_{\mu}-T^{\lambda}_{\lambda\mu}\right)\overline{\mathcal{T}}_{\nu\alpha}\right]=P^{\alpha}_{\beta}\left[T^{\lambda}_{\mu\alpha}n_{\lambda}\overline{\mathcal{E}}^{\mu} -\overline{\mathcal{P}}_{\mu}\nabla_{\alpha}v^{\mu}\right]+ \nb \\
    &-\frac{1}{\sqrt{\gamma}}\partial_\mu(\sqrt{\gamma}\cM^\m_a)V_\b^a\ .
\end{align}

\subsection{Energy and momentum in generalized Maxwell theory}

In the derivation of the Ward identities for diffeomorphisms, we used the variations with respect to the background sources. Applying those variations to the action \eqref{eq:F2}, we obtain the energy, the momentum and the stress currents: 
\begin{subequations}\label{eq:energymomentumcurved}
\ba
 \label{eq:ene_cur}
 \overline{\mathcal{E}}^{\mu} &=& \frac{1}{2}v^\mu \left( h^{\alpha\beta}v^\rho v^\sigma +\frac{\sigma}{2} h^{\alpha\beta} h^{\rho\sigma}\right)F_{\rho \alpha a} F_{\sigma\beta a}-\sigma \,h^{\mu\sigma}h^{\alpha\beta}v^\rho F_{\sigma\alpha a}F_{\rho\beta a} \nb \\ 
&\qquad &+\, \frac{4}{\sqrt{\g}}\p_{\s} G^{\s \n a}    v^{\r} \zerof{F}_{\r \l} \frac{\zeta P^{\mu}_{\nu}V^{\lambda}_a-(\beta+\zeta)P^{\lambda}_{\nu}V^{\mu}_a}{V^{2}} \ , \\ 
    \overline{\mathcal{P}}_{\mu} &=& -P^\alpha_\mu\left[ h^{\nu\sigma}v^{\rho}F_{\rho\sigma a}F_{\alpha\nu a}+\frac{2}{\sqrt{\gamma}}\partial_\sigma G^{\sigma \nu a}\,\beta n_\nu {\cal F}^{(0)}_{\alpha\lambda}\frac{V_a^\lambda}{V^2}\right] \ , \\
    \overline{\mathcal{T}}_{\mu\nu} &=& - \pq{v^{\a} v^{\b} \pr{P_{\m}^{\r} P^{\s}_{\n} - \frac{1}{2} h_{\m \n} h^{\r \s} } 
   - \frac{\s}{2} h^{\a \b} \pr{2 P_{\m}^{\r} P^{\s}_{\n} - \frac{1}{2} h_{\m \n} h^{\r \s} }} F_{\a \r a} F_{\b \s a} \nb \\ 
   &\qquad &-\,\frac{8}{\sqrt{\gamma}}\partial_\sigma G^{\sigma \alpha a}\left[(\beta+\zeta)\delta^\rho_{\s}-\zeta v^\rho n_{\s}\right] {\cal F}^{(0)}_{\rho\lambda}H_{\mu\nu a}^\lambda \ ,
\ea
\end{subequations}
where we have defined
\begin{equation}
    H_{\mu\nu a}^\lambda=\frac{1}{V^2}\left(P^\lambda_{(\mu}V_{\nu)}^b \delta_{ab}-\frac{1}{d} \frac{V_\mu^c V_\nu^d\delta_{cd}}{V^2}V_a^\lambda\right) \ .
\end{equation}
These quantities are explicitly gauge invariant. However, since the variation of the field strength ${\cal F}^{(0)}$ (see \eqref{defBmunu}) with respect to $V_\m^a$ is not gauge invariant, one might worry that some terms in the diffeomorphism Ward identities are not gauge invariant either. We show now that this is not the case, and the Ward identities for diffeomorphisms are actually gauge invariant.

The equation of motion for $a_\mu$ is 
\begin{equation}
    \label{eq:eom_a}
    \partial_\mu K^{\mu\nu}=0 \ ,\q \q \q K^{\mu\nu}=-K^{\nu\mu} \ ,
\end{equation}
where
\begin{equation}
    K^{\mu\nu}=8\partial_\sigma G^{\sigma\rho a} \left[(\beta+\zeta)\delta_\rho^{[\mu}V_a^{\nu]}-\zeta n_\rho v^{[\mu}V_a^{\nu]}\right] \ .
\end{equation}
On the other hand, the variation of the action with respect to $V_\mu^a$ contains gauge-invariant terms coming from the factors of $V^a_\mu$ that multiply the field strength $\cF^{(0)}$, plus a possibly gauge variant term due to the fact that the field strength ${\cal F}^{(0)}_{\mu\nu}$ contains $V^a_\mu$ (see \ref{defBmunu}),
\begin{equation}
\label{eq:pos_gau_var}
\delta S=-\int d^{d+1}x\, K^{\mu\nu}b_{\mu a}\delta V_\nu^a+\cdots \ ,
\end{equation}
where we have integrated by parts once. Comparing with \eqref{eq:genfuncvar}, we identify
\begin{equation}
\sqrt{\gamma}\cM_a^\mu =  K^{\mu\nu}b_{\nu a}+\cdots  \ .
\end{equation}
In the Ward identity for diffeomorphisms \eqref{eq:wardid}, we have the divergence of this quantity, which, recalling \eqref{eq:eom_a}, is gauge invariant on-shell:
\begin{equation}
\partial_\mu\left(\sqrt{\gamma}\cM_a^\mu\right) =  \frac{1}{2}K^{\mu\nu}\cF^{(1)}_{\mu\nu  a}+\cdots  \ .
\end{equation}

Let us add couplings to the monopole and dipole currents $J^\mu$ and $J^{\mu a}$, and study the gauge invariance of the diffeomorphisms Ward identity \eqref{eq:wardid} in this case. The monopole and dipole currents should change under a variation of $V_\mu^a$ as 
\begin{equation}
    \delta J^\mu=0 \ , \q \q \q  \delta J^{\mu a}=\frac{1}{\sqrt{\gamma}} N^{\mu\nu}\delta V_\nu^a \ ,
\end{equation}
such that 
\begin{equation}
    N^{\mu\nu}=-N^{\nu\mu}\ ,\q \q \q \partial_\mu N^{\mu\nu}=\sqrt{\gamma}J^\nu \ ,
\end{equation}
which, recalling that $\partial_{[\mu} V_{\nu]}^a = 0$, corresponds to compatibility with the monopole and dipole Ward identities \eqref{eq:curvedWardnoncovariant}.
When the currents are non-zero, the equation of motion for $a_\mu$ is modified into $\partial_\mu K^{\mu\nu}=\sqrt{\gamma}J^\nu$, and the variation of the action with respect to $V_\mu^a$ is
\begin{equation}
\delta S=\int d^{d+1}x\, \left[(K^{\mu\nu}-N^{\mu\nu}) b_{\nu a}+\sqrt{\gamma}Q^{\mu}_a\right]\delta V^a_{\mu} \ ,
\end{equation}
where $Q_a^\mu$ contains all the gauge-invariant contributions and reads
\begin{equation}\label{eq:Qamu}
    Q^\mu_a= \frac{4}{\sqrt{\gamma}}\left( \partial_{\sigma}G^{\sigma \nu b}\right)\left[(\beta+\zeta)\delta^\rho_{\nu}-\zeta v^\rho n_{\nu}\right]\frac{1}{V^2}\left( h^{\lambda \mu}\delta_{ba}-\frac{2}{d}\frac{V^{\lambda}_{b}V^{\mu}_a}{V^2}\right) \mathcal{F}^{(0)}_{\rho \lambda} \ .
\end{equation}
Comparing again with \eqref{eq:genfuncvar}, we have the following identification
\begin{equation}
\sqrt{\gamma}\cM_a^\mu=(K^{\mu\nu}-N^{\mu\nu})b_{\nu a}+\sqrt{\gamma}Q^{\mu}_a \ .
\end{equation}
Although $\cM_a^\mu$ is not gauge-invariant, its divergence is gauge invariant. Indeed, 
\begin{equation}
\partial_\mu\left(\sqrt{\gamma}\cM_a^\mu\right)= \sqrt{\gamma}\,\mathcal{K}^{\mu\nu}\cF^{(1)}_{\mu\nu  a} +\partial_\mu(\sqrt{\gamma} Q_a^\mu) \ ,
\end{equation}
where we have defined
\begin{equation}
    \sqrt{\gamma}\,\mathcal{K}^{\mu\nu}\equiv\frac{1}{2}\left(K^{\mu\nu}-N^{\mu\nu}\right) \ .
\end{equation}
Grouping the gauge-invariant terms of $\cM_a^\mu$ into $Q^\mu_a$, the Ward identities for the diffeomorphisms \eqref{eq::WardIdentitiesdiffeos} take the following gauge-invariant form
\begin{align}
    &n_{\alpha}\nabla_{\mu}\overline{\mathcal{E}}^{\mu}-\nabla_{\mu}\left( v^{\mu}\overline{\mathcal{P}}_{\alpha}\right)-h^{\mu\nu}\nabla_{\mu}\overline{\mathcal{T}}_{\nu\alpha}-T^{\lambda}_{\lambda\mu}\left( n_{\alpha}\overline{\mathcal{E}}^{\mu}-v^{\mu}\overline{\mathcal{P}}_{\alpha}-h^{\mu\nu}\overline{\mathcal{T}}_{\nu\alpha}\right)-\overline{\mathcal{P}}_{\mu}\nabla_{\alpha}v^{\mu}+ \nb \\
&+T^\l_{\m \a} n_\l \overline{ \mc{E}}^\mu-\mathcal{K}^{\mu\nu}\mathcal{F}_{\mu\nu a}^{(1)}V^a_{\alpha}- V^a_{\alpha}\left( \nabla_{\mu}-T^{\lambda}_{\lambda\mu}\right)Q^{\mu}_a = 0 \ .
\end{align}

The temporal and spatial projections of the diffeomorphism Ward identities yield
\begin{subequations}\label{eq:wardenergymomentum}
\begin{equation}\label{eq:wardenergy}
    \left( \nabla_{\mu}-T^{\lambda}_{\lambda\mu}\right)\overline{\mathcal{E}}^{\mu} =  -\Big[\overline{\mathcal{T}}_{\nu\alpha} \, h^{\mu(\nu}\nabla_{\mu}v^{\alpha)}+\overline{\mathcal{E}}^{\mu}\,T^{\lambda}_{\mu\alpha}n_{\lambda}v^{\alpha}\Big] \ , 
\end{equation}
\begin{equation}\label{eq:wardmomentum}
\begin{split}
    &\Big[\left( \nabla_{\mu}-T^{\lambda}_{\lambda\mu}\right)\left( v^{\mu} \overline{\mathcal{P}}_{\alpha}+v^\mu {\cal P}_\alpha^{\rm int}\right) +\left( \nabla_{\mu}-T^{\lambda}_{\lambda\mu}\right)(h^{\mu\nu}\overline{\mathcal{T}}_{\nu\alpha}+h^{\mu\nu}\mathcal{T}_{\nu\alpha}^{\rm int})\Big]P^{\alpha}_{\beta} =   \\ &\qquad \qquad \Big[\overline{\mathcal{E}}^{\mu}\,T^{\lambda}_{\mu\alpha}n_{\lambda} -\overline{\mathcal{P}}_{\mu} \, \nabla_{\alpha}v^{\mu} 
    +\left(v^\mu{\cal P}_\lambda^{\rm int} +h^{\mu\nu}\mathcal{T}_{\nu\lambda}^{\rm int} \right)\,\frac{V^\lambda_a\nabla_\mu V^a_{\alpha}}{V^2}\Big]P^\alpha_\beta -\mathcal{K}^{\mu\nu}\mathcal{F}_{\mu\nu a}^{(1)}V^a_{\beta}\ , 
\end{split}
\end{equation}
\end{subequations}
where we have defined an ``internal momentum" and an ``internal stress'' as follows
\begin{equation}\label{eq:internalmomentum}
    {\cal P}_\mu^{\rm int}=Q^\lambda_a  n_\lambda  V^a_\mu \ ,\qquad \mathcal{T}_{\mu\nu}^{\rm int}=h_{\mu\lambda} Q^\lambda_a V^a_\nu \ .
\end{equation}
We also used the fact that $Q^\mu_a$ can be expressed in terms of internal momentum and stress as
\begin{equation}
    Q^\mu_a=\left(v^\mu{\cal P}_\lambda^{\rm int} +h^{\mu\nu}\mathcal{T}_{\nu\lambda}^{\rm int} \right)\frac{V^\lambda_a}{V^2} \ .
\end{equation}

\section{Symmetric tensor theory in a curved background}\label{sec:pretko}

In the present section, we focus on the choice of parameters $\a= - \b=2 \zeta=1$ that yields Pretko's model. Then, we rewrite our curved-spacetime formulae in terms of a symmetric tensor field, analogously to what we did in  (\ref{eq:defsymmetricfieldinflatcase}) and (\ref{eq:fieldstrengthPretkoflat}) for the flat-spacetime case.  

In flat spacetime, the symmetric tensor theory arises by means of particular projections along the temporal and spatial directions of the dipole-invariant combination $b_{\m a} \d^{a}_{\n} - \p_{\m} a_{\n}$. In the curved case, in order to have covariance under diffeomorphisms, we need to replace the normal derivative with its covariant version.
We consider the combination
\be
\label{eq:defbetacombination}
\b_{\m \n} = b_{\m a} V^{a}_{\n} -  \dc_{\m}  a_{\n} + a_{\s} \frac{V^{\s}_{a} \dc_{\m} V_{\n} ^{a}}{V^{2}}  \ ,
\ee
where $V^{2}$ has been defined in (\ref{eq:defV2}). Such a combination is invariant under dipole transformations. Indeed,
\ba
\d \b_{\m \n} &=& \p_{\m} \l_{a}  V^{a}_{\n} -  \dc_{\m}\pr{ \l_{a} V_{\n}^{a}  }+  \l_{a}   \dc_{\m} V_{\n} ^{a} = 0 \ ,
\ea
where we have used $V^{\m}_{a} V^{b}_{\m} = V^{2} \d^{b}_{a}$.
Under monopole transformations, instead, $\beta_{\mu\nu}$ transforms as
\be
\label{eq:monopoletransfofbeta}
\d \b_{\m \n} = -  \dc_{\m}  \p_{\n} \l + \p_{\s} \l\frac{V^{\s}_{a} \dc_{\m} V_{\n} ^{a}}{V^{2}}  \ .
\ee
We denote by $A_{\mu\nu}$ the symmetric spatial part of the combination $\beta_{\mu\nu}$,
\ba
A_{\m \n} &\equiv& 2 P^{\r}_{\m} P^{\s}_{\n} \b_{(\r \s)} \ . 
\ea
In addition to $\beta_{\mu\nu}$, the theory also includes the dipole-invariant scalar $\phi=v^\mu a_\mu$. Its invariance follows from the condition \eqref{eq:closure_V}.

In curved spacetime, the field strength $F_{\mu\nu a}$ for the symmetric tensor reduces to
\begin{equation}
F_{\mu\nu a}=\cF^{(1)}_{\mu\nu a}-\cG_{\mu\nu a}^{(0)}+\frac{1}{2}\tilde \cG_{\mu\nu a}^{(0)}\ ,
\end{equation}
where, from the definitions given in (\ref{eq:defcurvaturesG}), we have
\ba
\cG_{\mu\nu a}^{(0)} &=& -2\partial_{[\mu} \left(\frac{V_a^\lambda}{V^2} \cF_{\nu]\lambda}^{(0)}\right) \ ,\\
\tilde \cG_{\mu\nu a}^{(0)}
&=& -2\partial_{[\mu} \left(\frac{V_a^\lambda}{V^2}P^\rho_{\ \nu]} \cF_{\rho\lambda}^{(0)}\right) \ .
\ea
The field strength $F_{\mu\nu a}$ can be written in terms of a vector potential,
\begin{equation}
F_{\mu\nu a}=2\partial_{[\mu} Z_{\nu] a} \ ,
\end{equation}
with
\begin{equation}
\label{eq:firstZexpression}
Z_{\mu a}= b_{\mu a}+ \frac{V_a^\lambda}{V^2} \cF_{\mu\lambda}^{(0)}-\frac{V_a^\lambda}{2 V^2} P^\rho_{\ \mu} \cF_{\rho\lambda}^{(0)} \ .
\end{equation}
The vector potential $Z_{\mu a}$ is invariant under monopole gauge transformations, and its variation under dipole gauge transformations is
\begin{equation}
\delta Z_{\mu a}= \partial_\mu \lambda_a \ .
\end{equation}
In Appendix \ref{app:gaugefieldpretko}, we derive the result
\begin{equation}
\label{eq:resultpotentialZ}
Z_{\mu a}= \partial_\mu \left(\frac{V_a^\lambda}{V^2} a_\lambda\right)+ \frac{ V_a^\lambda}{V^2}\left[ P^\rho_{\ \mu} \beta_{(\rho\lambda)} -n_\mu  \partial_\lambda\phi\right]+2 \phi\, n_{(\mu}\delta_{\alpha)}^\rho\Gamma^\alpha_{[\rho\lambda]}\frac{V_a^\lambda}{V^2} \ .
\end{equation}

In flat spacetime, the symmetric tensor gauge field of Pretko's model \eqref{eq:defsymmetricfieldinflatcase} is invariant under dipole gauge transformations. 
Starting from $Z_{\m a}$ in (\ref{eq:resultpotentialZ}), we can obtain a dipole-invariant vector potential without affecting the field strength $F_{\m \n a}$ simply by removing the total derivative term:
\ba
\label{eq:pretkogaugefieldcurved}
{\cal A}_{\mu a} &=& \left[P^\rho_{\ \mu} \beta_{(\rho\lambda)} -n_\mu  \partial_\lambda\phi+2\phi \, n_{(\mu}\delta_{\alpha)}^\rho\Gamma^\alpha_{[\rho\lambda]}\right]\frac{V_a^\lambda}{V^2} \nb \\
&=& \left[\frac{1}{2} A_{\mu\lambda} -n_\mu  \partial_\lambda\phi+ \phi\, n_{(\mu}\delta_{\alpha)}^\rho T^\alpha_{\rho\lambda}\right]\frac{V_a^\lambda}{V^2} \, .
\ea
We can deduce its transformation under monopole gauge transformation from the fact that it must be minus the transformation of the total derivative term that we have just removed, namely
\begin{equation}
\label{eq:gaugetransformPretkoPotential}
\delta {\cal A}_{\mu a}=-  \partial_\mu \left(\frac{V_a^\n}{V^2} \partial_\n \lambda\right) \, .
\end{equation}

We have thus shown that Pretko's model can be coupled to a curved (Aristotelian) spacetime consistently with gauge invariance and diffeomorfism covariance writing a Maxwell-like action in terms of the field strength $F_{\m \n a}= 2 \p_{[\m} {\cal A}_{\n]a}$ with the vector potential ${\cal A}_{\m a}$ given in (\ref{eq:pretkogaugefieldcurved}).

Let us examine the coupling of the dynamical fields to external currents. Using the decomposition
\be
a_{\m} = \phi\ n_{\m} + P_{\m}^{\s} a_{\s} \ , \q \q \q P^{\s}_{\m} = \frac{V^{\s}_{a} V^{a}_{\m}}{V^{2}} \ ,
\ee
and recalling the Ward identity (\ref{eq:curvedWardnoncovariant}), we get
\ba
S_{J} &=& \int d^{d+1} x\ \sqrt{\g} \pr{ a_{\m} J^{\m} + b_{\m a} J^{\m a}} \nb \\
&=& \int d^{d+1} x\ \sqrt{\g} \pr{\phi \, n_{\m} J^{\m} +  a_{\s} \frac{V^{\s}_{a} V_{\m}^{a}}{V^{2}} J^{\m} + b_{\m a} J^{\m a}} \nb \\
&=& \int d^{d+1} x\ \sqrt{\g} \pq{\phi\, n_{\m} J^{\m} +  a_{\s} \frac{V^{\s}_{a} }{V^{2}} \frac{1}{\sqrt{\g}} \p_{\m} \pr{\sqrt{\g} J^{\m a}} + b_{\m a} J^{\m a}} \nb \\
&=& \int d^{d+1} x\ \sqrt{\g} \pg{\phi\, n_{\m} J^{\m} + \pq{  b_{\m a} - \dc_{\m} \pr{ a_{\s} \frac{V^{\s}_{a} }{V^{2}}} }   J^{\m a}  } \ .
\ea
Finally, recalling (\ref{eq:defbetacombination}), we can write the action $S_J$ as
\ba
S_{J} =  \int d^{d+1} x \sqrt{\g} \pr{\phi\, n_{\m} J^{\m} +  \b_{\m \n} \frac{V_{a}^{\n}}{V^{2}}   J^{\m a}  }  \ .
\ea
Note that if the currents are the improved ones that satisfy (\ref{eq:Ward_improved}), the only components of $\beta_{\mu\nu}$ coupled to the currents are the symmetric spatial components $A_{\mu\nu}$.

\subsection{Energy and momentum conservation in flat space}
\label{sec:flatenergystress}

We obtained general expressions for the energy and momentum currents in a curved geometry in \eqref{eq:energymomentumcurved}, and identified an internal contribution to the diffeomorphism Ward identities \eqref{eq:Qamu}, \eqref{eq:internalmomentum}. From those expressions, fixing $\a= - \b=2 \zeta=1$ which connects to Pretko's theory, and taking the flat-spacetime limit, one obtains%
\footnote{We used the definition $\tilde{F}_{\mu\nu}=\partial_l F_{\mu\nu l}$ introduced in Subsection \ref{sec:mapflat} and the equations of motion for the symmetric tensor field \eqref{eq:eomssymm}.}
\begin{equation}
    \label{eq:com_zer}
    \overline{\mathcal{P}}_0=\overline{\mathcal{T}}_{0\mu}=Q_a^0=0 \ .
\end{equation}
The energy current is given by
\begin{equation}
   \overline{\mathcal{E}}^{0} = \frac{1}{2}\left(F_{0 k a} F_{0 k a}+\frac{\sigma}{2} F_{kl a} F_{kla}\right) \ ,\ \quad \overline{\mathcal{E}}^{i}= -\sigma F_{ik a}F_{0k a} \ , 
\end{equation}
while the momentum current is
\begin{equation}
\label{eq:mom_cur}
 \overline{\mathcal{P}}_i = \overline{\mathcal{P}}_i^S  
- \frac{1}{2} \tilde{F}_{0k} \zerof{F}_{ik}\ ;
\end{equation}
eventually, the stress tensor reads
\begin{equation}
\begin{split}
\label{eq:str_cur}
    \overline{\mathcal{T}}_{ij} =&\,   \overline{\mathcal{T}}_{ij}^S
    +\eta_{ij,kl}\left(-2 \tilde{F}_{0k}\zerof{F}_{0l} 
     +\frac{\sigma}{2} \tilde{F}_{nk}\zerof{F}_{nl}\right) \ ,
   \end{split}
\end{equation}
where we have introduced the symmetric and traceless shear tensor
\begin{equation}
    \eta_{ij,kl}\equiv\frac{1}{2}(\delta_{ik}\delta_{jl}+\delta_{il}\delta_{jk})-\frac{1}{d}\delta_{ij}\delta_{kl}\ .
\end{equation}
In \eqref{eq:mom_cur} and \eqref{eq:str_cur} we have denoted by $\overline{\mathcal{P}}_i^S$ and $\overline{\mathcal{T}}_{ij}^S$   the terms depending only on the symmetric tensor field, which explicitly are given by
\begin{equation}
    \label{eq:ter_sym}
    \overline{\mathcal{P}}_i^S = - F_{0ka} F_{ika},\quad \overline{\mathcal{T}}_{ij}^S =   -F_{0ia}F_{0ja}+\sigma F_{kia}F_{kja}+\frac{1}{2}\delta_{ij}\left(F_{0ka}F_{0ka}-\frac{\sigma}{2}F_{lka}F_{lka}\right)\ .
\end{equation}
The internal stress as introduced in \eqref{eq:internalmomentum} takes the following explicit form
\begin{equation}
    \mathcal{T}_{ij}^{\rm int} = \delta_{ik}Q_a^k\delta^a_j =\tilde{F}_{0j}\zerof{F}_{0i}-\frac{\sigma}{4}\tilde{F}_{kj}\mathcal{F}^{(0)}_{ki}-\frac{2}{d}\delta_{ij}\left(\tilde{F}_{0k}\zerof{F}_{0k} -\frac{\sigma}{4} \tilde{F}_{kl}\mathcal{F}^{(0)}_{kl} \right) \ .
\end{equation}
On the other hand, note that ${\cal P}_\mu^{\rm int}$ introduced in \eqref{eq:internalmomentum} vanishes because, according to \eqref{eq:com_zer}, we have $Q^0_a=0$.

The terms \eqref{eq:ter_sym}, which only depend on the symmetric tensor, coincide with the usual energy-momentum tensor for a relativistic (Lorentz-invariant) theory, if $\sigma$ is identified with the speed of light squared and the index $a$ is treated as an internal index. The rest of the terms depend on $b_{ta}$ and $b_{[ij]}$ through $\zerof{F}$. Note that the diffeomorphism Ward identity  \eqref{eq:wardenergymomentum} in flat space also depends on these components through $\onef{F}$. Using the fact that
\begin{equation}
    K_{0i}=-K_{i0}=\tilde{F}_{0i} \ , \q \q \q K_{ij}=-\frac{\sigma}{2}\tilde{F}_{ij}\ ,
\end{equation}
the momentum Ward identity is
\begin{equation}
    \partial_0  \overline{\mathcal{P}}_i+\partial_k (\overline{\mathcal{T}}_{ki}+\mathcal{T}_{ki}^{\rm int})=-\tilde{F}_{0k}\onef{F}_{0k i}+\frac{\sigma}{4}\tilde{F}_{kl}\onef{F}_{kl i}\ .
\end{equation}
In principle this could lead to an additional constraint for the time and antisymmetric components of $b_{\mu a}$, but we have checked that the Ward identity is satisfied identically just using the equations of motion for the symmetric components \eqref{eq:eomssymm}. Thus, we can simplify the diffeomorphism Ward identity to an expression involving only the symmetric field
\begin{equation}
    \partial_0  \overline{\mathcal{P}}_i^S+\partial_k \overline{\mathcal{T}}_{ki}^S=-\tilde{F}_{0k}F_{0k i}+\frac{\sigma}{4}\tilde{F}_{kl}F_{kl i}\ .
\end{equation}
In either case, there is no conserved momentum current. There is an associated conserved current that one can construct by writing the right hand side of the diffeomorphism Ward identity as a total derivative. However, the resulting momentum and stress currents are not gauge invariant.

In contrast to the momentum Ward identity, the Ward identity for the energy \eqref{eq:wardenergy} in flat space implies that the energy is both gauge-invariant and conserved
\begin{equation}
    \partial_0 \overline{\mathcal{E}}^{0} +\partial_i \overline{\mathcal{E}}^{i} =0\ .
\end{equation}
This can be checked explicitly using the equations of motion \eqref{eq:eomssymm}.

\section{Discussion}\label{sec:discuss}

Using the formulation of dipole symmetry in terms of ordinary gauge fields, we have shown that Pretko's symmetric tensor theory and other related generalized Maxwell theories can be consistently coupled to a curved background, in a covariant and gauge-invariant way. This opens the doors to similar generalizations of higher-multipole theories, for instance a generalization to the traceless case, with a conserved second moment of the charge, would be straightforward using the gauge fields introduced in \cite{Caddeo:2022ibe}.

By examining gauge transformations in a curved geometry, we have been able to determine the conditions for the existence of a global dipole symmetry. Using these conditions, we have identified a family of geometries with global dipole symmetry, namely a Cartesian product of time with a conformally flat spatial manifold. The conditions are similar, although not exactly the same, as those found to covariantly couple the symmetric tensor gauge field to curved space without spoiling the gauge invariance \cite{Slagle:2018kqf,Bidussi:2021nmp,Jain:2021ibh,Armas:2023ouk}. In our construction, contrary to the tensor gauge formulation, the gauge invariance is preserved even when a global dipole symmetry does not exist.

Utilizing the coupling to a curved background we have derived the Ward identities for diffeomorphisms, encoding energy and momentum conservation, and we have obtained explicit expressions for the energy and momentum currents in flat space. The Ward identities imply that there is no gauge-invariant and conserved momentum current, in agreement with previous analysis \cite{Jain:2021ibh}. Despite this, our construction shows that a non-conserved and gauge-invariant momentum current is still compatible with a covariant formulation of the theory and its coupling to a curved background. There is, however, an obstruction to promote the metric to a dynamical field preserving diffeomorphisms: the non-zero background field for internal translations $V_\mu^a$ explicitly breaks the would-be dynamical gauge symmetry.

In $d=2$ spatial dimensions the theory of the symmetric tensor field can be interpreted as a particle-vortex dual of elasticity, with the momentum of the elastic medium being \cite{Pretko_2018}
\begin{equation}
    \mc{P}_a^{\rm dual}=\frac{1}{2}\epsilon_{ij}\epsilon_{ab}F_{ijb},\qquad  \mc{T}_{ia}^{\rm dual}=\epsilon_{ik}\epsilon_{ab}F_{0kb}\ .
\end{equation}
The dual momentum is conserved by virtue of the Bianchi identity in \eqref{eq:eomssymm}.
From this point of view there is a notion of conserved momentum, but the coupling to a dynamical metric would require working in the dual formulation, or perhaps dualizing the metric as in \cite{Tsaloukidis:2023bvz}. Also, this does not extend to higher dimensions. The dual momentum can be seen as the current of a 0-form symmetry carrying an internal index, while in larger $d$ the associated dual symmetry would be a $(d-2)$-form symmetry instead. For instance, in $d=3$ we have a conserved dual 1-form symmetry with current
\begin{equation}
    \mc{J}_{ia}^{\rm dual}=\frac{1}{2}\epsilon_{ikl}F_{kla},\qquad  \mc{J}_{ija}^{\rm dual}=\epsilon_{ijk}F_{0ka}\ .
\end{equation}
Summarizing, we confirm that the symmetric tensor gauge theory cannot be coupled to dynamical gravity, even though it can be consistently introduced in an arbitrary curved background geometry. This is an unusual feature since these two aspects typically go hand in hand in less exotic theories.

\section*{\large Acknowledgments}

We would like to thank Akash Jain and Stefan Prohazka for useful discussions. The work of E.A. is supported by the Severo Ochoa fellowship PA-23-BP22-170. This work is partially supported by the Spanish Agencia Estatal de Investigaci\'on and the Ministerio de Ciencias, Innovaci\'on y Universidades through the Spanish grant PID2021-123021NB-I00. 
A.C. acknowledges funding by Germany’s Excellence Strategy through the Würzburg-Dresden
Cluster of Excellence on Complexity and Topology in Quantum Matter - ct.qmat (EXC 2147, project-id
390858490), and by the Deutsche Forschungsgemeinschaft (DFG) through the Collaborative Research
centre “ToCoTronics”, Project-ID 258499086—SFB 1170.
A.C. would like to thank the IBS Science Culture Center of Daejeon for support during the workshop ``Effective Field Theory Beyond Ordinary Symmetries''. 

\appendix

\section{Newton-Cartan geometries}\label{app:NewtonCartanreview}

In this appendix, we review the basics of Newton-Cartan geometry and fix the notation used in the main text.
The Newton-Cartan geometry features a set of tensors $\pg{n_\m, h^{\m \n}, v^{\m}}$, where $n_\m$ is a clock one-form, $h^{\m \n}$ the (degenerate) inverse spatial metric and $v^\m$ a velocity field. These objects satisfy the following relations,
\be
\label{newcartanidentities}
v^\m n_\m = 1 \ , \q \q \q n_\m h^{\m \n} = 0  \ , \q \q \q h^{\m \n} h_{\n \r} = \d^\m _\r - v^\m n_\r \equiv P^\m _\r \ ,
\ee
the last of which can be taken as defining $h_{\m \n}$. $P^\m _\n$ plays the role of a spatial projector:
\be
n_\m P^\m _\n =  P^\m _\n  v^\n = 0 \ .
\ee
Once $v^{\m}$ is introduced, one can construct an invertible tensor $\g_{\m \n}$,
\be
\label{appdef:gammametric}
\g_{\m \n} = n_{\m}n_{\n} + h_{\m \n} \ , \q \q \g^{\m \n} = v^{\m} v^{\n} + h^{\m \n} \ , \q \q \g^{\m \n} \g_{\n \r} = \d^{\m}_\r \ ,
\ee
that allows defining the volume form in our $(d+1)$-dimensional manifold $\mc{M}$,
\be
\text{vol} \pr{\mc{M}} = \frac{1}{(d+1)!} \varepsilon _{\m_{0} \dots \m_{d}} \dd x^{\m_{0}} \wedge \dots \wedge \dd x^{\m_{d}} \ , \q \q \q \varepsilon _{\m_{1} \dots \m_{d}}  \equiv \sqrt{\g} \epsilon _{\m_{1} \dots \m_{d}}  \ .
\ee
Here, $\g \equiv \text{det}\g_{\m \n}$,  $\varepsilon$ is the Levi-Civita tensor and $\e$ the Levi-Civita symbol $\e_{0 \dots d} =1$. From these, we define
\be
\varepsilon ^{\m_{0} \dots \m_{d}} = \g^{\m_{0} \m_{0}'} \dots \g^{\m_{d} \m_{d}'} \varepsilon _{\m_{0}' \dots \m_{d}'} =  \g^{-1} \varepsilon _{\m_{0} \dots \m_{d}} \ .
\ee
This allows defining the Hodge dual operation in the usual way.

We can introduce a connection to covariantly differentiate tensors. In the Newton-Cartan context, it is possible to choose a connection such that $n_{\m}$ and $h^{\m \n}$ are covariantly constant; however, in the general case where $\p_{[\m} n_{\n]} \neq 0$, such a connection displays (at least) temporal torsion. We define
\be
\label{eqapp:NCconnection}
\dc_{\m} X^{\n} = \p_{\m} X^{\n} + \G^{\n}_{ \m \l} X^{\l} \ , \q \q \q \dc_{\m} Y_{\n} = \p_{\m} Y_{\n} - \G^{\l} _{ \m \n} Y_{\l} \ ,
\ee
with the connection given by\footnote{The most general connection with the properties specified above also includes a term of the form $h^{\m \s} n_{(\n} F_{\r ) \s}$, where $F_{\m \n}$ is a two-form that is usually identified with the field strength of the $U(1)$ connection $A_{\m}$ that couples to the mass current. As we deal with systems without Galilean boost invariance, we can set $A_{\m}=0$.}
\be
\G^{\m}_{\n \r} = v^{\m} \p_{\n} n_{\r} + \frac{1}{2} h^{\m \l} \pr{\p_{\n} h_{\l \r} + \p_{\r} h_{\l \n} - \p_{\l} h_{\n \r}} \ .
\ee
In this case, the torsion is only temporal and reads
\be
T^{\m}_{\n \r} = 2 v^{\m} \p_{[\n} n_{\r]} \ .
\ee
Finally, given a generic vector field $Z^{\m}$, the following formula turns out to be useful when it comes to integration by parts:
\be
\label{eqapp:covariantdivergence}
\pr{\dc_{\m} - T_{\n \m} ^{\n}} Z^{\m} = \frac{1}{\sqrt{\g}} \p_{\m} \pr{\sqrt{\g} Z^{\m}} \ .
\ee

\section{Gauge potential of symmetric tensor theory}
\label{app:gaugefieldpretko}

In this appendix, we show the computation that leads from the definition (\ref{eq:firstZexpression}) to the result (\ref{eq:resultpotentialZ}) presented and used in the main text.

For convenience, let us rewrite here (\ref{eq:firstZexpression}), the starting expression of $Z_{\m a}$,
\begin{equation}
Z_{\mu a}= b_{\mu a}+ \frac{V_a^\lambda}{V^2} \cF_{\mu\lambda}^{(0)}-\frac{V_a^\lambda}{2 V^2} P^\rho_{\ \mu} \cF_{\rho\lambda}^{(0)} \ .
\end{equation}
Recalling that in components $\zerof{F}_{\m \n}$ is defined as
\begin{equation}
\mc{F}^{(0)}_{\m \n} = 2 \p_{[\m} a_{\nu]} - b_{\mu a}V_\nu^{a} + b_{\nu a} V_\mu^{a} \ ,
\end{equation}
we can reorganize the terms in $Z_{\mu a}$ in the following way:
\begin{equation}
\label{eq:expressionWoriginalfields}
Z_{\mu a}=\frac{ V_a^\lambda}{V^2}\left[\frac{1}{2}P^\rho_{\ \mu} \left(b_{\rho b}V^b_\lambda-\partial_\rho a_\lambda+b_{\lambda b} V^b_\rho-\partial_\lambda a_\rho\right) +\partial_\mu a_\lambda-v^\rho n_\mu \partial_\lambda a_\rho\right] \ ,
\end{equation}
where we have used that $P^\rho_{\ \mu}V_\rho^a=V_\mu^a$. Now we will introduce the dipole-invariant quantities
\be
\b_{\m \n} = b_{\m a} V^{a}_{\n} -  \dc_{\m}  a_{\n} + \frac{ a_\sigma}{V^{2}} V^{\s}_{a} \dc_{\m} V_{\n} ^{a} \ , \q \q \q \phi=v^\alpha a_\alpha \ .
\ee
We will replace
\begin{equation}
b_{\mu a}V^a_\nu-\partial_\mu a_\nu=\beta_{\mu\nu}+a_\alpha \pr{ -\Gamma_{\mu\nu}^\alpha- 
\frac{V^{\a}_{a} \dc_{\m} V_{\n} ^{a} }{V^{2}} }  \ ,
\end{equation}
and use
\begin{equation}
v^\rho n_\mu \partial_\lambda a_\rho=n_\mu \left[\partial_\lambda \phi-a_\rho\partial_\lambda v^\rho\right] \ .
\end{equation}
We will also write
\begin{equation}
\frac{V_a^\lambda}{V^2}\partial_\mu a_\lambda=\partial_\mu \left(\frac{V_a^\lambda}{V^2} a_\lambda\right)-\partial_\mu\left(\frac{V_a^\lambda}{V^2}\right) a_\lambda \ .
\end{equation}
We get
\begin{equation}
Z_{\mu a}=  \partial_\mu \left(\frac{V_a^\lambda}{V^2} a_\lambda\right)+  \frac{  V_a^\lambda}{V^2}\left[P^\rho_{\ \mu} \beta_{(\rho\lambda)} -n_\mu \partial_\lambda\phi\right]+a_\alpha S^\alpha_{\mu a} \ , 
\end{equation}
where
\begin{equation}
S^\alpha_{\mu a}=- \partial_\mu\left(\frac{V_a^\alpha}{V^2}\right) +\frac{V_a^\lambda}{V^2}\pq{ 
P^\rho_{\ \mu}\pr{ -\Gamma_{(\rho\lambda)}^\alpha  - \frac{V^{\a}_{a} \dc_{(\r} V_{\l )} ^{a}  }{V^{2}} }
+n_{\mu}\partial_{\lambda} v^\alpha } \ .
\end{equation}
If we project,
\be
V_\alpha^b S^\alpha_{\mu a}=\frac{ V^\lambda_a}{V^2}\pq{
\partial_\mu V_\lambda^b+ P^\rho_{\ \mu}\pr{ -\Gamma_{(\rho \lambda)}^\alpha V_\alpha^b -  \dc_{(\r} V_{\l )} ^{b}  }
+n_{\mu}\partial_{\lambda} v^\alpha V_\alpha^b } \ .
\ee
Now we use $\partial_\mu V_\lambda^b=n_\mu v^\rho \partial_\rho V_\lambda^b+P^\rho_{\ \mu}\partial_\rho V_\lambda^b$ and
\ba
V_a^\lambda P^\rho_{\ \mu}\partial_\rho V_\lambda^b 
&=& V_a^\lambda P^\rho_{\ \mu}\partial_{(\rho} V_{\lambda)}^b=V_a^\lambda P^\rho_{\ \mu}\left[\nabla_{(\rho} V_{\lambda)}^b+\Gamma_{(\rho\lambda)}^\alpha V_\alpha^b\right]  \ .
\ea
Then,
\begin{equation}
V_\alpha^b S^\alpha_{\mu a}=\frac{V^\lambda_a}{V^2}\left[n_\mu v^\rho\partial_\rho V_\lambda^b+n_\mu V_\rho^b \partial_\lambda v^\rho \right]=  \frac{V^\lambda_a}{V^2} n_\mu \partial_\lambda(v^\rho V_\rho^b)=0  \ ,
\end{equation}
which is consistent with the gauge transformation $\delta W_{\mu a}=\partial_\mu \lambda_a$. We now split
\begin{equation}
a_\alpha=n_\alpha v^\sigma a_\sigma+ P^\sigma_{\ \alpha} a_\sigma=n_\alpha\phi+ a_\sigma\frac{V^\sigma_b}{V^2} V_\alpha^b \ .
\end{equation}
Then, we can write
\begin{equation}
Z_{\mu a}= \partial_\mu \left(\frac{V_a^\lambda}{V^2} a_\lambda\right)+ \frac{ V_a^\lambda}{V^2}\left[ P^\rho_{\ \mu} \beta_{(\rho\lambda)} -n_\mu  \partial_\lambda\phi\right]+\phi \,n_\alpha S^\alpha_{\mu a} \ .
\end{equation}
We will use
\begin{equation}
-\partial_\mu\left(\frac{V_a^\alpha}{V^2}\right) n_\alpha=-\frac{1}{V^2}\partial_\mu V^\alpha_a n_\alpha=\frac{V_a^\lambda}{V^2} \partial_\mu n_\lambda \ .
\end{equation}
This gives
\ba
n_\alpha\, S^\alpha_{\mu a}
&=&\frac{V_a^\lambda}{V^2}\pq{ \partial_\mu n_\lambda   - P^\rho_{\ \mu}  \Gamma_{(\rho\lambda)}^\alpha n_\alpha + n_{\mu}\partial_{\lambda} v^\alpha n_\alpha  } \nb \\
&=&\frac{V_a^\lambda}{V^2}\left[\partial_\mu n_\lambda-P^\rho_{\ \mu}\partial_{(\rho}n_{\lambda)}-n_{\mu} v^\rho \partial_{\lambda} n_\rho \right]\nb \\
&=& 2n_{(\mu}\delta_{\alpha)}^\rho\Gamma^\alpha_{[\rho\lambda]}\frac{V_a^\lambda}{V^2} \ ,
\ea
where we have used $\Gamma^\alpha_{\mu\nu}n_\alpha=\partial_\mu n_\nu$ and $\Gamma^\lambda_{[\mu\nu]}=v^\alpha \partial_{[\mu}n_{\nu]}$. Finally, the gauge field written in covariant form is
\begin{equation}
Z_{\mu a}= \partial_\mu \left(\frac{V_a^\lambda}{V^2} a_\lambda\right)+ \frac{ V_a^\lambda}{V^2}\left[P^\rho_{\ \mu} \beta_{(\rho\lambda)} -n_\mu  \partial_\lambda\phi\right]+2 \phi\, n_{(\mu}\delta_{\alpha)}^\rho\Gamma^\alpha_{[\rho\lambda]}\frac{V_a^\lambda}{V^2} \ .
\end{equation}

\bibliographystyle{utphys}
\bibliography{bibliography}

\end{document}